\documentclass{mn2e}

\usepackage{amsfonts}
\usepackage{amsmath}
\usepackage{graphicx}

\title[Marked correlations in galaxy formation models]
{Marked correlations in galaxy formation models}

\author[R. K. Sheth, A. J. Connolly \& R. Skibba]
       {Ravi K. Sheth$^{1}$, Andrew J. Connolly$^2$ 
        \& Ramin Skibba$^2$\thanks{E-mail:   shethrk@physics.upenn.edu (RKS); 
                          ajc@phyast.pitt.edu (AJC);
                          ramin@phyast.pitt.edu (RS)}\\
$^{1}$Department of Physics \& Astronomy, University of Pennsylvania, 
      209 S. 33rd Street, Philadelphia, PA 19130, USA\\
$^{2}$Department of Physics \& Astronomy, University of Pittsburgh, 
      3941 O'Hara Street, Pittsburgh, PA 15260, USA}

\newcommand{\bm}[1]{{\mbox{\boldmath $#1$}}}

\def\fun#1#2{\lower3.6pt\vbox{\baselineskip0pt\lineskip.9pt
        \ialign{$\mathsurround=0pt#1\hfill##\hfil$\crcr#2\crcr\sim\crcr}}}

\begin{document}

\pagerange{\pageref{firstpage}--\pageref{lastpage}}


\maketitle

\label{firstpage}

\begin{abstract}
The two-point correlation function has been the standard statistic 
for quantifying how galaxies are clustered.  The statistic uses 
the positions of galaxies, but not their properties.  Clustering as 
a function of galaxy property, be it type, luminosity, color, etc., 
is usually studied by analysing a subset of the full population, 
the galaxies in the subset chosen because they have a similar range 
of properties.  
We explore an alternative technique---marked correlations---in which 
one weights galaxies by some property or `mark' when measuring 
clustering statistics.  Marked correlations are particularly 
well-suited to quantifying how the properties of galaxies correlate 
with their environment.  Therefore, measurements of marked 
statistics, with luminosity, stellar mass, color, star-formation 
rate, etc. as the mark, permit sensitive tests of galaxy formation 
models.  We make measurements of such marked statistics in 
semi-analytic galaxy formation models to illustrate their utility.  
These measurements show that close pairs of galaxies are expected 
to be red, to have larger stellar masses, and to have smaller star 
formation rates.  
We also show that the simplest unbiased estimator of the particular 
marked statistic we use extensively is very simple to measure---it 
does not require construction of a random catalog---and provide an 
estimate of its variance.  
Large wide-field surveys of the sky are revolutionizing our view of 
galaxies and how they evolve.  Our results indicate that application 
of marked statistics to this high quantity of high-quality data 
will provide a wealth of information about galaxy formation.   
\end{abstract}



\section{Introduction}
It has been thirty five years since Totsuji \& Kihara (1969) published 
their measurement of the galaxy correlation function.  
Since then, measurements in ever larger catalogs have shown their 
initial estimate of $\xi(r)\propto r^{-1.8}$ on scales smaller than 
about ten Megaparsecs was accurate (e.g., Maddox et al. 1990; 
Percival et al. 2001; Hamilton \& Tegmark 2002; Connolly et al. 2002).  
Recent work in the 2dFGRS (Colless et al. 2001) and SDSS 
(Abazajian et al. 2005) surveys has shown definitively that 
clustering depends on galaxy type, color and luminosity 
(Norberg et al. 2002; Zehavi et al. 2002, 2005).  
In what follows, we will speak of any combination of properties which 
distinguish galaxies from one another as marks.  

Marks can take discrete or continuous values, and they may be scalars 
or vectors.  The traditional morphological type of a galaxy is a 
discrete scalar mark.  Luminosity, color, X-ray hardness, AGN 
activity and star formation rate are all examples of continuous scalar 
marks.  A typical neural network classifies morphological types as a 
list of probabilities ${\bm p}=(p_1,\cdots,p_n)$,  where $p_i$ is the 
probability a galaxy is of type $T_i$ (e.g., 
Storrie-Lombardi et al. 1992; Lahav et al. 1996).  
If one classified galaxies using this probability vector (rather 
than simply using the type $T_i$ which had the largest $p_i$), then 
morphological type is a vector mark.  Principal component analyses of 
galaxy spectra output a vector ${\bm c}=(c_1,\cdots,c_n)$, where 
$c_i$ denotes the contribution of eigenspectrum $i$ to the object's 
spectrum (e.g. Connolly et al. 1995).  
The spectral classification ${\bm c}$ can, therefore, be used as a 
vector mark.  

In most analyses to date (e.g., Hamilton 1988; 
Mo, B\"orner \& Zhou 1989; Willmer, da Costa \& Pellegrini 1998; 
Benoist et al. 1999; Norberg et al. 2002; Zehavi et al. 2005), 
the parent galaxy catalog is cut into subsamples based on the mark 
(most often morphological type or luminosity), and the clustering 
in the subsample is studied by treating each galaxy in it equally.  
That is to say, the mark is used to define the subsample, but is not 
considered further.  
This is equivalent to relabeling all marks to be `zeros' or `ones', 
depending on whether or not the value of the mark crossed a threshold 
value, and then computing the correlation function of the galaxies 
marked as `ones'.  There are two reasons why this procedure is not 
ideal.  First, the choice of critical threshold is somewhat arbitrary, 
particularly in the case of marks like luminosity and color which take 
a continuous range of values.  Second, throwing away the actual value 
of the mark represents a loss of information. 
Therefore, one might well ask:  Why not weight each galaxy by its 
actual mark, rather than relabelling to ones and zeros?  

With the advent of large wide-field space-- and ground--based surveys 
of the sky, many of the marks mentioned above can now be measured with 
sufficient precision that contamination by measurement error is not a 
serious concern.  Indeed, many of these marks have been, or can be, 
reliably measured in a number of datasets now available.  
For example, the luminosities in the 2MASS survey are expected to be 
indicators of total stellar mass; 
the ROSAT all sky survey can be used to estimate the hardness of the 
spectra of X-ray sources; 
the GALEX mission will provide estimates of star formation rates out 
to redshifts of order unity.  
The time is ripe to take advantage of the additional information 
provided by weighting each galaxy by its mark.  

Marked correlations in the SSRS2, IRAS 1.2 Jy and PSCz surveys, 
using luminosity and/or type as mark were measured by 
Beisbart \& Kerscher (2000).  The marked correlations showed 
evidence that the luminosities of close pairs of galaxies were 
larger than the mean on separations larger than about $10h^{-1}$Mpc.  
Beisbart \& Kerscher did not compare their measurements with 
specific model predictions, so their results have had less 
impact on galaxy formation models than they might otherwise have 
had.  

Semi-analytic galaxy formation models (White \& Rees 1979; 
White \& Frenk 1991; Kauffmann et al. 1999; Somerville \& Primack 1999; 
Cole et al. 2000; Springel et al. 2005) provide the basis for 
our current understanding of how and why different galaxy types 
are differently biased tracers of the dark matter field.  
They also provide a useful framework for seeing if simple models are 
consistent with the observed correlations between luminosity and 
velocity dispersion or circular velocity (Faber \& Jackson 1976; 
Tully \& Fisher 1977), morphology and density (Dressler 1980), 
morphology or density and star formation rate (Lewis et al. 2002; 
Gomez et al. 2003; Kauffmann et al. 2004), etc.  

Although the predictions of these models depend on a number of free 
parameters, and they do not always provide perfect descriptions of the 
data, they have proven to be useful for gaining insight into how 
changes in the physics of galaxy formation affect the properties of 
the galaxy population.  These models output a number of marks which 
are potentially observable.  These include luminosity, size, velocity 
dispersion, morphology, star-formation rate, local density, and so on, 
and how these marks evolve.  
The main goal of the present paper is to measure various 
marked correlation functions in these models, so as to illustrate 
how marked statistics can provide direct physical insight into the 
processes of galaxy formation.  

Section~\ref{define} introduces the family of pairwise marked correlation 
functions, and defines the particular generalization of the usual 
correlation function we will use.  
Section~\ref{sams} shows measurements of various marked correlation 
functions made in semi-analytic models of galaxy formation.  
Section~\ref{rescale} discusses the effects of rescaling marks prior 
to measuring marked statistics.  Such rescaling is necessary if one 
wishes to compare galaxy formation models with observations, if, 
as is often the case, the distribution of marks in the models is not 
the same as in the data.  
A final section summarizes our findings, and discusses some 
extensions.  
An Appendix provides a brief discussion of the 
bias and variance of our estimators of marked statistics (our 
estimators are unbiased, but they are not minimum variance).  
In essence, the estimator we use throughout this paper is particularly 
straightforward to implement because it does not require construction 
of a random catalog; in this respect, marked statistics are 
substantially easier to estimate than the usual unweighted 
correlation functions.

\section{Marked statistics}\label{define}
Marked correlations measure the clustering of marks.  
Since the positions at which the marks are measured may themselves 
be clustered, marked statistics are defined in a way which accounts 
for this.  This is the subject of Sections~\ref{moments} and~\ref{norm}.  
Marked statistics are usually studied to see if marks depend on 
environment.  Section~\ref{correlations} shows that marked statistics 
have another very useful application which has not been emphasized 
previously.  Namely, different marks frequently correlate with each 
other:  marked statistics can be used to quantify if the correlations 
between marks depend on environment.  

\subsection{Definitions}\label{moments}
For example, let $\bar\rho$ denote the mean density of particles, 
and let $\bar m$ denote the mean mark, averaged over all particles.  
Now consider a particle with mark larger than this mean value.  
Are the particles neighbouring it also likely to have larger marks?  
One way to quantify how likely this is is to compute the ratio of the 
mean mark to $\bar m$ of pairs of particles as a function of pair 
separation.  The typical number of pairs at separation $r$ is 
$\bar\rho^2 [1 + \xi(r)]$, where $\xi$ is the two point correlation 
function.  Therefore, the mean mark is 
\begin{eqnarray}
 {\cal M}_1(r) &=& {\sum  [m({\bm x}) + m({\bm y})]\,
                          {\cal I}(|{\bm x}-{\bm y}|-r) 
                  \over 2\bar m\ \sum {\cal I}(|{\bm x}-{\bm y}|-r)} 
                  \nonumber\\
               &=& {\sum  [m({\bm x}) + m({\bm y})]\,
                          {\cal I}(|{\bm x}-{\bm y}|-r) 
                  \over 2\bar m\ \bar\rho^2 [1+\xi(r)]} ,
\end{eqnarray}
where ${\cal I}(x)=0$ unless $x=0$, and the sum is over all galaxy 
pairs.  We have divided by $\bar m$, so ${\cal M}_1(r)=1$ for all $r$ 
if there are no correlations between marks.  
Analogously, the $n$th-order mark is defined by 
\begin{equation}
 {\cal M}_n(r) =  {\sum [m({\bm x}) + m({\bm y})]^n\,
                          {\cal I}(|{\bm x}-{\bm y}|-r) 
                   \over (2\bar m)^n\ \bar\rho^2 [1+\xi(r)]}.
 \label{Mn}
\end{equation}
In what follows, we will concentrate on the mean square mark 
\begin{displaymath}
 {\cal M}_2(r) =  {\sum [m^2({\bm x}) + m^2({\bm y}) + 2m({\bm x})m({\bm y})]\,
                          {\cal I}(|{\bm x}-{\bm y}|-r) 
                   \over 4\bar m^2\ \bar\rho^2 [1+\xi(r)]}.
\end{displaymath}
If all the marks are the same, then ${\cal M}_2=1$.  If the marks are 
independent, with mean $m_1=\bar m$ and mean square $m_2$, then 
${\cal M}_2 = (m_2 + m_1^2)/2\bar m^2 = (m_2/\bar m^2 + 1)/2$.  
To see why this is sensible, compute the variance: 
 ${\cal V}ar(r) = {\cal M}_2(r) - {\cal M}_1(r)^2 
                = (m_2/\bar m^2 - 1)/2$.  
This shows that ${\cal V}ar$ equals half the variance of the individual 
marks, as it should (recall it is the variance of one half times 
the sum of two independent marks).  

Two types of terms contribute to ${\cal M}_2$.  The term which 
involves a product of two marks is proportional to 
\begin{equation}
 M(r)\equiv {\sum m({\bm x})m({\bm y})\,{\cal I}(|{\bm x}-{\bm y}|-r) 
 \over \bar m^2\,\sum {\cal I}(|{\bm x}-{\bm y}|-r) } 
 \equiv {1+W(r)\over 1+\xi(r)} ,
 \label{weighted}
\end{equation}
where we have defined $W(r)$ for the following reason.  
The only difference between the sums in the numerator and denominator 
is that, in the numerator, the $i$th particle contributes a weight 
$m_i/\bar m$, whereas in the denominator, the weight is unity for 
all particles.  Since the denominator is one plus the usual unweighted 
correlation function, the numerator can be thought of as being one plus 
a weighted correlation function.  In effect, the denominator divides-out 
the contribution to the weighted correlation function which comes from 
the spatial distribution of the points, leaving only the contribution 
from the fluctuations of the marks.  

\subsection{Estimators and the weighted correlation function}
There are a number of reasons why this ratio, $M(r)$, is the measure 
of marked correlations on which we will concentrate.  First, fast 
estimators of unweighted correlations (which account for edge effecs, 
etc.) have been available for some time (e.g., Peebles 1980).  If we 
write the usual estimator for the unweighted correlation function as 
$DD/RR$, then the weighted correlation function is $WW/RR$, and the 
marked statistic above is simply $WW/DD$.  This has two important 
implications.  First, allowing for weights requires only a simple 
modification of algorithms which estimate unweighted correlations 
(e.g. Moore et al. 2000).  Second, because edge effects appear in 
both the numerator and denominator, this marked correlation ratio is 
less sensitive to edge effects than is, e.g., the $DD/RR$ 
estimator of the unweighted correlation function 
(e.g. Beisbart \& Kerscher 2000).  
Perhaps more importantly, note that the $RR$ factor cancels out, so 
that the marked statistic can be estimated without constructing a 
random catalog; this significantly reduces the computational burden 
for estimating the statistic.  
Third, errors on the measurement are readily estimated, as 
described in Appendix~\ref{estimator}.  (It is relatively common 
to approximate the errors by remaking the measurement after 
randomizing the marks, and repeating a large number of times.  
We show that this is a reasonable approximation on the small scales 
where we find the most interesting results, but leads to an 
underestimate of the true errors on larger scales.)  
And finally, it is by thinking of marked correlations as this 
ratio of weighted to unweighted correlations that one is able to 
construct theoretical models of marked correlations 
(Sheth, Abbas \& Skibba 2004; Sheth 2005).  

\subsection{Measurement errors}
If the mark is not perfectly measured, so the observed mark $o_i$ of 
object $i$ is $o_i = m_i + e_i$, where $e_i$ is the error in the 
measurement, then the measured marked correlation function will be 
$M_o(r) = \langle(m_i+e_i)(m_j+e_j)|r\rangle$ , where the notation 
denotes the average over all pairs $i$ and $j$ separated by $r$.  
Thus, $M_o(r) = M_m(r)+M_e(r)$ plus a term which allows for the fact 
that the mark of one object is correlated with the error made in 
measuring the marks of the other objects.  If there is no such 
correlation, and if the mean measurement error for objects does 
not depend on position, and if the errors themselves are uncorrelated, 
then $M_o(r) = M_m(r)$:  
in this case, the measured marked correlation will be an unbiased 
representation of the true correlation.  Of course, the precision with 
which this marked correlation function can be determined will depend on 
the amplitudes of the $e_i$s.  Appendix~\ref{estimator} provides a 
more detailed discussion.  

\subsection{Normalization}\label{norm}
The expressions above follow the conventions in the statistical 
literature (e.g. Stoyan 1984; Stoyan \& Stoyan 1994), and so 
normalize the marks by the mean mark.  There is no apriori reason 
for this choice.  For instance, we could have chosen to normalize 
by the rms value instead.  In models where large scale correlations 
are expected to be small, our convention makes the large scale limit 
$M(r)\to 1$, which is intuitively easy to understand.  Normalizing 
by the square root of the mean squared mark instead makes the 
amplitude of the small-scale signal easier to understand, since 
this convention would have $M(r\to 0)\to 1$, but then the large 
scale value would depend on the ratio of the mean and rms values 
of the marks.  This convention might be useful if one is interested 
in studying marked statistics in a density field which was obtained 
by smoothing the mark-weighted point distribution.  

Moreover, the expressions above implicitly assume that $\bar m$ is 
not zero.  If marks are not positive definite, then $\bar m\ne 0$ 
is not guaranteed, and one may well have to normalize by the rms value.  
To avoid this complication, in what follows, we will only consider 
marks which are positive definite.  
Analysis in Section~\ref{rescale} suggests that monotonic rescalings 
of the marks will not seriously compromise their use.  
Thus, for instance, when we use galaxy colors (which are proportional 
to the log of the ratio of the luminosity in two different wavebands) 
as marks, we use the ratio of the luminosities, rather than the log 
of this ratio.  

\subsection{Cross-correlations}\label{cross}
So far, we have implicitly assumed that the mark used for both particles 
was the same.  There is no reason why we could not have used different 
marks for the two particles of each pair.  For example, we could use 
effective radius and local density as two marks to study if galaxies 
in denser regions are smaller or larger than average.  
Marked cross-correlations could also be used to study if star formation 
rates in galaxies are affected by AGN activity in their neighbors.  
To estimate cross correlations of this type, the expressions given 
previously remain true, with the replacements 
$m({\bm x})m({\bm y})/\bar m^2\to 
 m_a({\bm x}) m_b({\bm y})/\bar m_a\bar m_b$, where the subscripts 
$a$ and $b$ denote the two different types of marks.  

\subsection{Correlations between marks}\label{correlations}
Many galaxy observables correlate with each other:  e.g., the 
velocity dispersion of a galaxy correlates with its size, luminosity, 
and color.  Marked statistics provide a simple way to see if such 
correlations depend on environment.  Note that this differs from 
the previous subsection which described marked cross-correlations 
between a mark of one galaxy and another mark of its neighbour; 
here we are interested in quantifying how the correlation between 
two (or more) marks of the same galaxy depends on the surrounding 
environment.  

Suppose each galaxy has two weights, $m$ and $w$, and that a 
plot of the two versus one another (each galaxy is a point in this 
two-dimensional plot) shows a correlation, but there is scatter around 
the mean correlation.  Let $w = f(m)$ denote this mean correlation.  
A question which often arises is:  Does the scatter correlate with 
environment?  
If we write $w_i = f_i + e_i$, where $f_i\equiv f(m_i)$, then 
\begin{eqnarray*}
 M_w(r) &=&\left\langle{w_i\over\bar w}{w_j\over\bar w}\Big\vert r\right\rangle
 = \left\langle {[f_i + e_i]\over\bar f}{[f_j + e_j]\over\bar f}\Big\vert r\right\rangle \nonumber\\
 &=& \left\langle {f_i f_j + f_i e_j + f_j e_i + e_ie_j\over \bar f^2}\Big\vert r\right\rangle, 
\end{eqnarray*}  
where the notation is intended to show that the averages are over all 
pairs with separation $r$.  Evidently, $M_w$ is a (suitably weighted) 
combination of $M_f$, $M_e$, and the cross-correlation between $f$ and $e$.  
If the amount $e_i$ by which galaxy $i$ scatters from the mean $w-m$ 
relation is independent of the marks $m_j$ of the other galaxies, then 
$\langle e_i f_j|r\rangle = \langle e_i|r\rangle\langle f_j|r\rangle$
and $\langle e_i e_j|r\rangle = \langle e_i|r\rangle\langle e_j|r\rangle$.  
If the scatter is independent of environment, then 
$\langle e_i|r\rangle = 0$ for all $r$.  
This shows that, if there is no correlation in the scatter, then 
 $M_w(r)\to M_{f(m)}(r)$.
Hence, one can determine if the correlation between marks depends on 
scale by comparing $M_w(r)$ with $M_{f(m)}(r)$.  
If $M_m(r)$ depends on scale, then $M_w=M_m$ only if $f(m)\propto m$.  
If $f(m)$ is a more complicated function of $m$, then $M_w$ differs 
from $M_m$ by a scale dependent factor even if $f(m)$ is independent 
of $r$.  


\subsection{An illustrative example}\label{haloexclusion}

\begin{figure}
 \centering 
 \vspace{-1cm}
 \includegraphics[width=1.1\hsize]{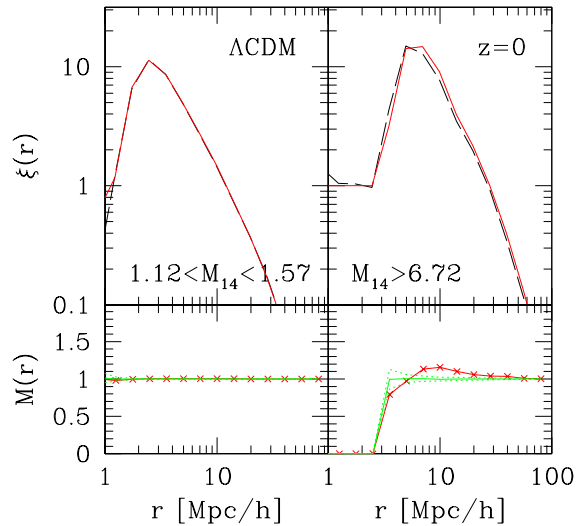}
 \vspace{-1cm}
\caption{Correlation functions of massive halos in numerical 
simulations of a $\Lambda$CDM cosmology.  Top panels show the 
number (dashed) and mass (solid) weighted correlation functions, 
and panels on the bottom show marked correlations (crosses).  The 
dotted curves in the bottom panels show the mean plus and minus the 
standard deviation of measurements made after randomizing the marks. }
\label{xihalos}
\end{figure}

Figure~\ref{xihalos} shows an example of what marked correlation 
functions measure.  
The underlying point process was the halo distribution in the 
Hubble Volume simulation (Evrard et al. 2002) of a flat $\Lambda$CDM 
cosmology ($\Omega_0=0.3,h=0.7,\sigma_8=0.9$).  
The dashed lines in the top panels show the unweighted correlation 
function of halos with masses in the range 
$1.12\times 10^{14}M_\odot/h \le M\le 1.57\times 10^{14}M_\odot/h$ 
(left, about $2\times 10^5$ halos) and greater than 
$6.72\times 10^{14}M_\odot/h$ (right, $2.5\times 10^4$ halos).  
The solid lines in the top panels show the result of weighting each 
halo by the ratio of its mass to the mean halo mass in the subsample 
when computing the correlation function.  Comparison of the two panels 
shows that on large scales, the more massive halos are more strongly 
clustered.  The decrease at small $r$ is a consequence of volume 
exclusion:  halos do not overlap.  The more massive halos occupy larger 
volumes, so volume exclusion matters (i.e., the correlation functions 
turn over) at larger scales in the panel on the right 
(this effect was discussed by Mo \& White 1996, and quantified by 
Sheth \& Lemson 1999).  

Let $C(r)$ and $W(r)$ denote the dashed and solid curves in the top 
panels.  The symbols with solid lines drawn through them in the bottom 
panels show the marked correlations $M(r) = [1+W(r)]/[1+C(r)]$, where 
the mark is the halo mass.  There is no trend in the panel on the 
left, because the range of halo masses in it is small.  As a result, 
almost all halos have the same mass, and hence the same mark.  
In contrast, the symbols in the panel on the right show a hump at 
10$h^{-1}$Mpc, indicating that the most massive halos tend to have 
separations of about that scale.  

One of the virtues of marked correlations is that there is a simple 
way to assess the statistical significance of such a feature.  
The approximately horizontal curves bounded by dotted lines in the 
bottom panels show the mean and the standard deviation around the 
mean when the same measurement is made after randomizing the marks.  
(The curves show results averaged over one hundred random 
realizations.)  Comparison of the symbols with these randomized 
curves shows that the hump at 10$h^{-1}$Mpc in the panel on the 
right is statistically significant.  (Appendix~A provides a more 
careful discussion of the uncertainty on the measured statistic, 
and shows that this randomization procedure slightly mis-estimates 
the true scatter, since it assumes there are no correlations between 
the marks.  However, since the correlations between marks we find 
are not very different from unity, this estimate based on 
randomization is not far from the true answer.)  

Although $M(r)\approx 1$ at large scales, it drops below unity on 
scales smaller than $5h^{-1}$Mpc.  This indicates that the halos which 
contribute pairs at small separations are the least massive subset of 
the population, as one would expect from volume exclusion.  
Notice that the marked correlations illustrated this effect without 
having to separate the catalog up into small bins in mass.  
Therefore, they allow a measurement of what is in this case a 
`mass segregation' effect which is of greater statistical significance 
than it would be in any one of the smaller catalogs.  

\begin{figure*}
\centering 
 \vspace{-2.5cm}
 \includegraphics[width=1.1\hsize]{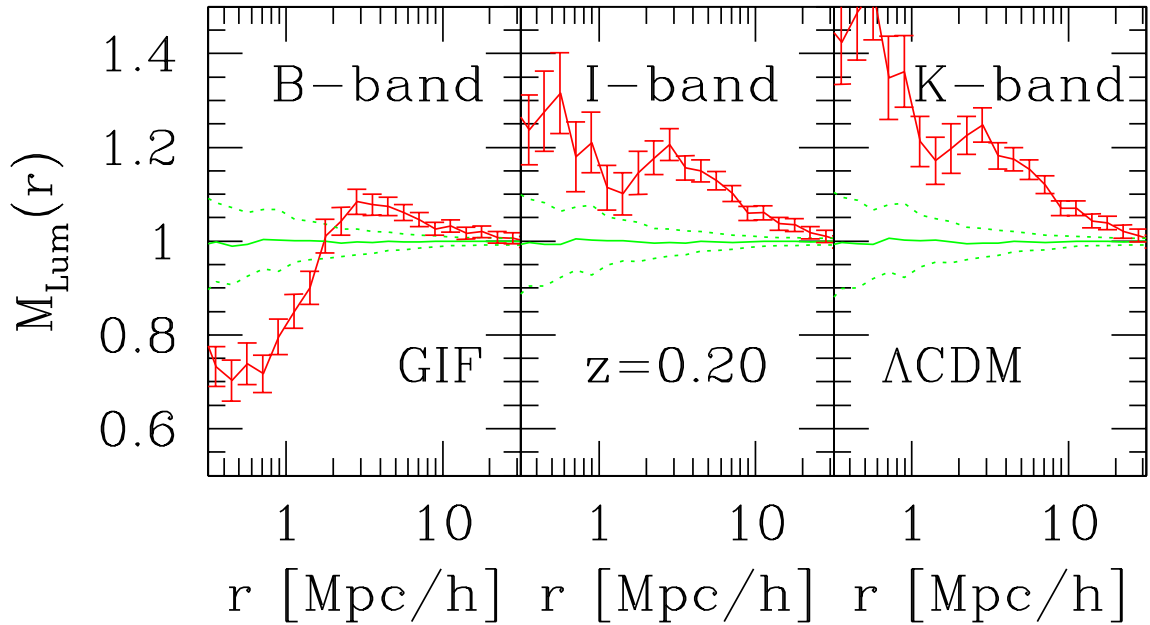}
 \vspace{-8.5cm}
 \caption{Marked correlations in the semi-analytic models with 
  $B$, $I$ and $K$-band luminosity at $z=0.20$ as the mark.  
  In each panel, jagged solid line shows the marked correlation, 
  and error bars show the estimated uncertainty on the measurement 
  derived in Appendix~A.  For comparison, horizontal line shows the 
  mean obtained by averaging over one hundred realizations of the 
  same measurement after randomizing the marks, and dotted lines 
  show the standard deviation around this mean.  When luminosity is 
  the mark, the marked correlation is a strong function of wavelength. }
 \label{xilum}
\end{figure*}

\section{Predictions of semi-analytic galaxy formation models}\label{sams}
It is straightforward to make similar measurements, using various 
marks, in semi-analytic galaxy catalogs.  
In what follows, we will use the galaxy formation models of 
Kauffmann et al. (1999) to illustrate the sort of information that 
marked correlations provide.  This is not because we feel other models 
are significantly worse, but because these model galaxy catalogs are 
available to the public ({\tt www.mpa-garching.mpg.de/Virgo}).  
The models populate the dark matter distribution of the GIF 
simulation with galaxies:  the background cosmology is a flat 
$\Lambda$CDM universe with $(\Omega,h,\sigma_8) = (0.3,0.7,0.9)$, 
and the simulation box is a cubical comoving volume $141h^{-1}$Mpc 
on a side.  All distances we show below are comoving.  

We show measurements of marked statistics in these semi-analytic 
galaxy catalogs for subsamples of galaxies which contain more than 
$2\times 10^{10} h^{-1}M_\odot$ in stars.  Such measurements are best 
thought of as being related to measurements in volume-limited, rather 
than magnitude-limited galaxy catalogs.  
At $z=0.2$ (this redshift was chosen to approximately match the 
median redshifts of the Two Degree Field Galaxy Redshift Survey and 
the Sloan Digital Sky Survey; c.f. Colless et al. 2001 and 
York et al. 2000) the semi-analytic galaxy catalog contains 
14,665 objects.  This number is about thirty percent smaller at 
$z=1.05$.  

\subsection{Luminosity}
Figure~\ref{xilum} shows marked correlations in which luminosity was 
used as the mark.  The various panels show measurements made 
in the semi-analytic models at $z=0.2$, at various wavelengths.  
In each panel, the jagged lines show the measurement, and error 
bars are estimated following the analysis in Appendix~A.  
The figure shows clear evidence that the shape of the marked correlation 
function depends on wavelength.  In all cases, $M_{\rm Lum}(r)$ 
falls gently at separations larger than about 3~Mpc, but the small 
scale behaviour is very different.  
It is interesting that this 3~Mpc scale is similar to that at which 
$\xi(r)$ shows a feature:  this is the scale on which the statistic 
becomes dominated by pairs in separate halos.  

\begin{figure}
\centering 
 \vspace{-1cm}
 \includegraphics[width=1.2\hsize]{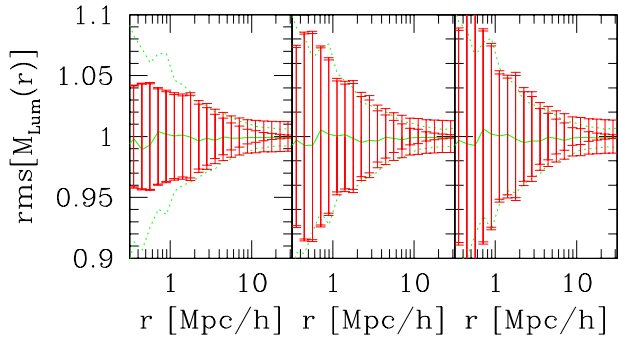}
 \vspace{-4cm}
 \caption{Comparison of various estimates of the uncertainty on the 
          measurement; the three panels show the same set of marks 
          as in Figure~\ref{xilum}.  
          Larger error bars show the full estimate from 
          Appendix~A; smaller error bars show the result of ignoring 
          the contribution from equation~(\ref{nonPoisson}).  
          The standard deviation of the marked statistic measured in 
          many (one hundred) realizations of the point distribution 
          with randomized marks is shown by the dotted curves.}
 \label{errors}
\end{figure}

Comparison of the curves in the different panels shows that while all 
bands are reasonably similar on larger scales, the different bands are 
quite different from each other on smaller scales.  
In contrast to the redder band-passes, the marked correlation decreases 
with scale for the $B-$band.  Evidently, the blue-band luminosities of 
galaxies with near neighbours (i.e., closer than a Mpc or so) are smaller 
than they are on average.  Since the $B-$band luminosity is more 
sensitive to recent star formation than the $I-$ or $K-$bands, it is 
interesting to study what happens when star-formation rate is used as 
the mark.  

We will make two additional points before we do so.  
First, although we have not shown this, we have found that if we 
restrict attention to any one band, then the amplitude of $M(r)$, 
especially on small scales, depends on the range of luminosities in 
the sample.  In general, increasing the range of luminosities makes 
the small-scale dependence of $M(r)$ on $r$ more dramatic.  
The reason for this was hinted at in Section~\ref{norm}:  if close 
pairs tend to have larger marks, then the amplitude of $M(r)$ 
depends on the range of luminosities in the sample.  

Second, consider the horizontal solid line at $M\sim 1$ on all scales 
in each panel.  This line shows the mean value of the statistic 
averaged over one hundred realizations of the catalog after 
randomizing the marks.  The dotted lines show the standard deviation 
around this mean, and this standard deviation is sometimes used as 
an estimate of the error on the measurement of $M(r)$.  
Notice that the difference between the solid and dotted lines is 
similar to the size of the error bars on the measurement, indicating 
that this randomized-marks procedure of estimating errors does not 
grossly mis-estimate the true errors, at least on small scales.  

Figure~\ref{errors} provides a more direct comparison.  
Dotted curves show the error estimate from randomizing the marks.  
The two sets of error bars show the two contributions to the error 
estimate from Appendix~A; the smaller estimate comes from ignoring 
the contribution from equation~(\ref{nonPoisson}).  
On small scales, the dotted lines and full error bars are in good 
agreement, except in the $B$-band.  In $B$-, $M(r)$ itself is smaller 
than unity, so the error estimate from Appendix~A is also smaller 
than the one obtained from randomizing the marks.  On larger scales, 
the estimate from Appendix~A becomes significantly larger than the 
one from randomizing the marks.  Since $M(r)$ itself is unity on 
these scales, this difference strongly suggests that our estimator, 
$WW/DD$, can be improved---while unbiased, is not minimum variance.  
This hypothesis is supported by the fact that, in this regime, the 
error estimate is dominated by the term in equation~(\ref{nonPoisson}).  

\subsection{Colors and star-formation}
\begin{figure}
\centering 
 \vspace{-1cm}
 \includegraphics[width=1.2\hsize]{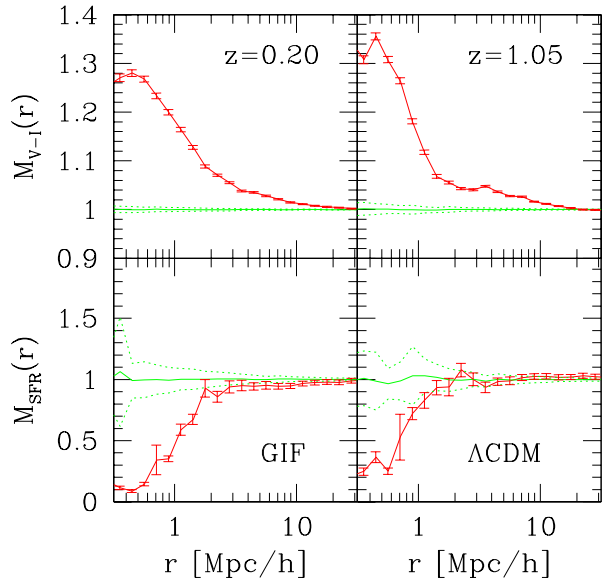}
 \vspace{-1cm}
 \caption{Marked correlations at $z=0.2$ and $z=1.05$ in the GIF 
 semi-analytic galaxy formation models (Kauffmann et al. 1999).  
 Top and bottom panels show results when the mark is restframe 
 $V-I$ color and star formation rate, respectively.  Curves with 
error bars show the marked correlations with associated uncertainties.  
Horizontal lines show the mean of one hundred realizations of the 
same measurement after randomizing the marks, and dotted lines show 
the standard deviation around this mean. }
 \label{xisfr}
\end{figure}

Figure~\ref{xisfr} shows the marked correlations of the same set of 
objects when color and star formation rate are used as marks (curves 
with error bars).  
As in the previous figure, the solid and dotted curves show the 
mean (averaged over one hundred realizations) and the standard 
deviation around the mean when the same measurement is made after 
randomizing the marks.  
The error estimates indicate that, at separations smaller than a 
Mpc or so, star formation rates are substantially lower than the 
mean, whereas the $V-I$ colors of close pairs are larger 
(i.e., redder) than the mean.  

Comparison of the panels on the left with those on the right shows 
how the marked correlations evolve.  
Notice that the two marked correlations depend very differently on 
scale, suggesting that color and star-formation rate are biased very 
differently relative to the underlying dark matter distribution.  
On the other hand, the scale on which $M_{V-I}(r)$ suddenly increases 
is similar to the scale on which $M_{\rm SFR}(r)$ decreases, suggesting 
that similar physics gives rise to both effects.  

Why does this happen?  Many close pairs come from galaxies in groups 
and clusters.  In the models, such galaxies started forming their 
stars at higher redshifts than average.  By the present time they are 
likely to have turned all the gas available to them into stars.  
Therefore, at low redshifts (e.g, $z<1$), the star formation rates in 
cluster galaxies are low,  and their colors are red.  
Figure~\ref{xisfr} shows that the marked correlation functions are 
sensitive to these effects.  Since the top panel is weighted 
by $V-I$, and redder colors mean larger values of $V-I$, the 
marked correlations in the top panels show a dramatic increase on 
scales smaller than the typical virial radius of a large cluster.  
Since star formation within clusters has essentially switched off, 
the trend in the bottom is the opposite.  
Thus, in the models, it is no accident that the scale on which 
$M_{V-I}$ suddenly increases is similar to the scale on which 
$M_{\rm SFR}$ decreases.  

Notice that the scale at which the two curves suddenly change is 
smaller at higher redshifts, and that the shift in scale is easily 
detected in both cases.  This is not unexpected in models where 
star-formation is lowest in the most massive halos.  
Comparison with similar measurements in the data will show if these 
predictions of the models are correct.  
If not, the gastrophysics which has been used to model galaxy 
formation should be refined.  
For instance, GALEX observes galaxies out to redshifts of order 
unity, so it is well suited to performing this test.  

\subsection{Stellar mass and the mass-to-light ratio}
We were prompted to study the star-formation rate because of the 
differences between the various panels of Figure~\ref{xilum}.  
The discussion above suggests that the drop on scales smaller 
than an Mpc in the $B$-band is consistent with Figure~\ref{xisfr} 
which showed a decrease in the star formation rate on small scales.  
Luminosities in the redder bands are less sensitive to recent star 
formation, so they are expected to be better tracers of the total 
stellar mass.  

\begin{figure}
\centering 
 \vspace{-4.25cm}
 \includegraphics[width=1.2\hsize]{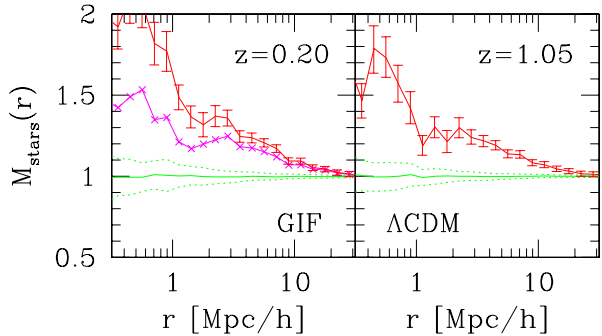}
 \vspace{-1cm}
\caption{Marked correlations in which stellar mass is the mark 
 (solid lines with error bars).  For comparison, crosses in the panel 
 on the left show what happens when $K-$band luminosity is the mark.}  
\label{ximstar}
\end{figure}

To illustrate, the curves with error bars in Figure~\ref{ximstar} 
show marked correlations in the semi-analytic models at $z=0.2$ and 
$z=1.05$ when the stellar mass is used as mark.  Comparison of the 
two panels shows that $M(r)$ evolves:  at both redshifts, close pairs 
tend to more massive, but the scale on which this is a significant 
effect is smaller at high redshift.  Moreover, the amplitude of $M(r)$ 
also evolves:  stellar masses of close pairs at high redshift were not 
as different from the mean as they are today.  The shift in amplitude 
and in scale means that similar measurements in data should provide 
sharp constraints on the models.  

The crosses in the panel on the left show the result of using the 
$K-$band luminosity as the mark (for clarity, we have not shown 
the associated error estimates, since these are shown in 
Figure~\ref{xilum}).  Notice that the same bumps and wiggles 
are present for both $M_K$ and $M_{stars}$.  
This suggests that the two marks, $K$-band luminosity and stellar 
mass, do indeed reflect similar physics.  
The discussion in Section~\ref{correlations} indicates that 
comparison of these two marked correlations can be used to study 
if the correlation between stellar mass and $L_K$ is the same in 
all environments.  In particular, if $L_K$ is linearly proportional to 
stellar mass, then the two marked correlations will be the same only 
if this correlation is independent of environment.  
(Specifically, if $L_1 = a M_1 + e_1$ with 
 $\langle L\rangle = a\langle M\rangle$, then the marked statistic is 
 $\langle L_1L_2|r\rangle/a^2\langle M\rangle^2 
 = \langle M_1M_2|r\rangle/\langle M\rangle^2 
         + 2\langle M_1e_2|r\rangle/a\langle M\rangle^2
         + \langle e_1 e_2|r\rangle/a^2\langle M\rangle^2$;
the final two terms vanish only if the correlation is independent of 
environment.)   

\begin{figure}
\centering 
 \includegraphics[width=0.9\hsize]{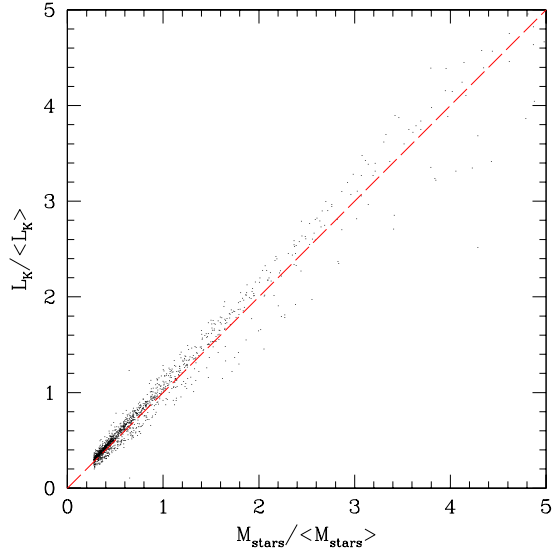}
\caption{Correlation between $K$-band luminosity and stellar mass.  
Since this correlation is linear, the two marked correlations will 
be the same only if there are no environmental effects.  }  
\label{m2l}
\end{figure}

Figure~\ref{m2l} shows that $L_K\propto M_{\rm stars}$ (we have 
normalized each by their respective mean values across the population).  
Hence, the difference between the two curves in the left panel of 
Figure~\ref{ximstar} suggests that the correlation between mass 
and light does depend on environment:  close pairs tend to have 
larger mass-to-light ratios, suggesting that the mass-to-light 
ratio is larger in dense regions.

\section{The effects of re-scaling}\label{rescale}
The semi-analytic models studied here do not match the observed 
distribution of galaxy properties precisely.  For instance, 
these models do not match the SDSS luminosity function.  
On the other hand, they are reasonably successful at describing 
the luminosity dependence of clustering.  
For this reason, we envision that constraints on such models will 
take the following form.  
If the models successfully describe the luminosity dependence of 
clustering, then they have successfully identified the halos in which 
galaxies form.  If marked correlations are not changed in any 
essential way if the absolute values of the weights are altered 
without changing the rank ordering, then marked statistics provide 
a way of determining if, in addition to correctly identifying the 
halos in which galaxies form, the models also determine the 
correct rank ordering of galaxy properties within halos.  

For such a test to work, we must understand the effect that 
monotonic rescalings of the marks have on marked statistics.  
With this in mind, consider the following toy model.  
All galaxies come in either of two types of groups.  
Groups of one type contain $m_1$ galaxies, each having weight $W_1$, 
and the density run of galaxies around centres the group centers 
follows a Gaussian distribution with scale length $s_1\propto m_1^{1/3}$.  
Galaxies in groups of the second type follow a Gaussian distribution 
with scale length $s_2$, have weights $W_2$, and there are 
$m_2\propto s_2^3$ galaxies per group.  
Specifically, the density run around each type of group is 
$\rho(r)/\bar\rho = \Delta_{\rm nl}\, \exp(-r^2/2/s_i^2)/\sqrt{2\pi}$, 
so $m_i = \Delta_{\rm nl}\bar\rho\,2\pi s_i^3$.  

\begin{figure}
\centering 
 \includegraphics[width=0.9\hsize]{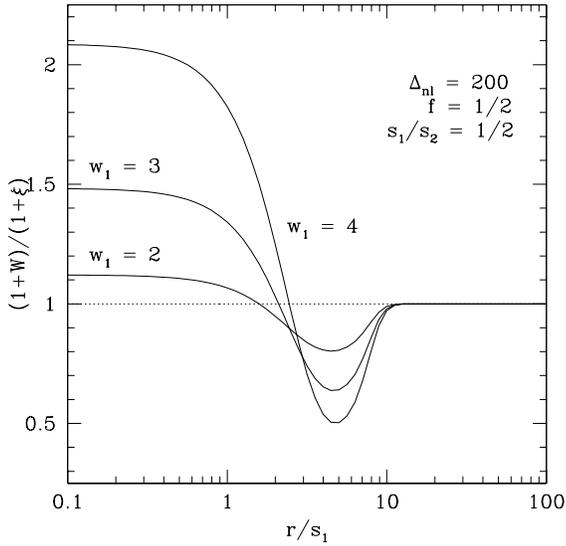}
 \caption{Marked correlation function for the model described in the text.  
          Note that the complex scale dependence of the statistic is 
          the same for a range of values of $w_1$, suggesting that 
          although rescaling marks may change the statistic 
          quantitatively, it does not result in qualitative changes.  }
 \label{gmarks}
\end{figure}

If $f$ denotes the fraction of groups of type 1, then the mean 
weight is $\bar W = [fm_1W_1 + (1-f)m_2W_2]/[fm_1 + (1-f)m_2]$.  
Hence, the normalized weights are $w_1 = W_1/\bar W$ and 
$w_2 = W_2/\bar W = 1 + (1-w_1)(s_1/s_2)^3\,f/(1-f)$.  
In this model, 
\begin{eqnarray}
 {1+W\over 1+\xi} 
  = {1 + kf w_1^2\, {\rm e}^{-r^2/4s_1^2} 
      + k(1-f) (s_2/s_1)^3 w_2^2\, {\rm e}^{-r^2/4s_2^2}\over 
   1 + kf\, {\rm e}^{-r^2/4s_1^2} + k(1-f)(s_2/s_1)^3\, {\rm e}^{-r^2/4s_2^2}}
\end{eqnarray}
where $k = [\Delta_{\rm nl}/4\sqrt{\pi}]/[f + (1-f)(s_2/s_1)^3]$.  
Rescaling the marks corresponds to changing $W_1$; the fact that the 
marks are normalized then determines the rescaled value of $W_2$.  
Thus, given $f$ and $s_1/s_2$, the dependence of the expression above 
on $w_1$ gives the range of possible shapes that the marked statistic 
can possibly take.  Figure~\ref{gmarks} illustrates when $w_1>1$.  
In this case, the marked statistic shows complex scale dependence, 
but this dependence is qualitatively similar for the different 
choices of $w_1$.  (We have not shown $w_1<1$; this case has 
$(1+W)/(1+\xi)$ greater than unity on all scales, with a maximum on 
intermediate scales.  In this respect, the scale dependence is less 
complex than that shown in Figure~\ref{gmarks}, but again, it is 
qualitatively similar for all $w<1$.)

\begin{figure}
\centering 
 \vspace{-1cm}
 \includegraphics[width=\hsize]{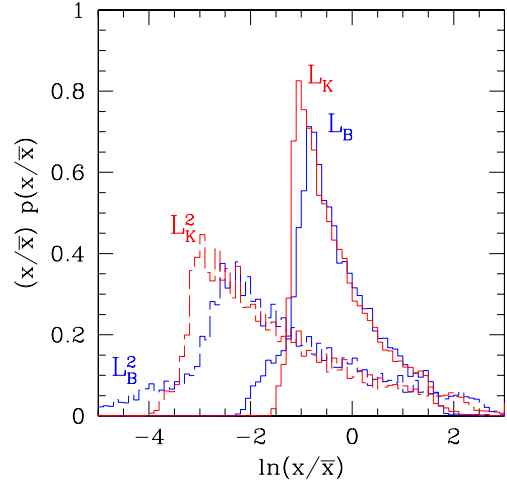}
 \vspace{-1cm}
 \caption{Distribution of normalized marks, $x/\bar x$, where 
 $\bar x$ is the mean mark for $x=L_B$, $L_B^2$, $L_K$ and $L_K^2$.  
 Notice that the distributions of normalized $L_B$ and $L_K$ are rather 
 similar; nevertheless, the associated marked statistics shown in 
 Figures~\ref{xilum} and~\ref{rescaleBIK} are very different.}  
\label{pdfBK}
\end{figure}

\begin{figure*}
\centering 
 \vspace{-2cm}
 \includegraphics[width=\hsize]{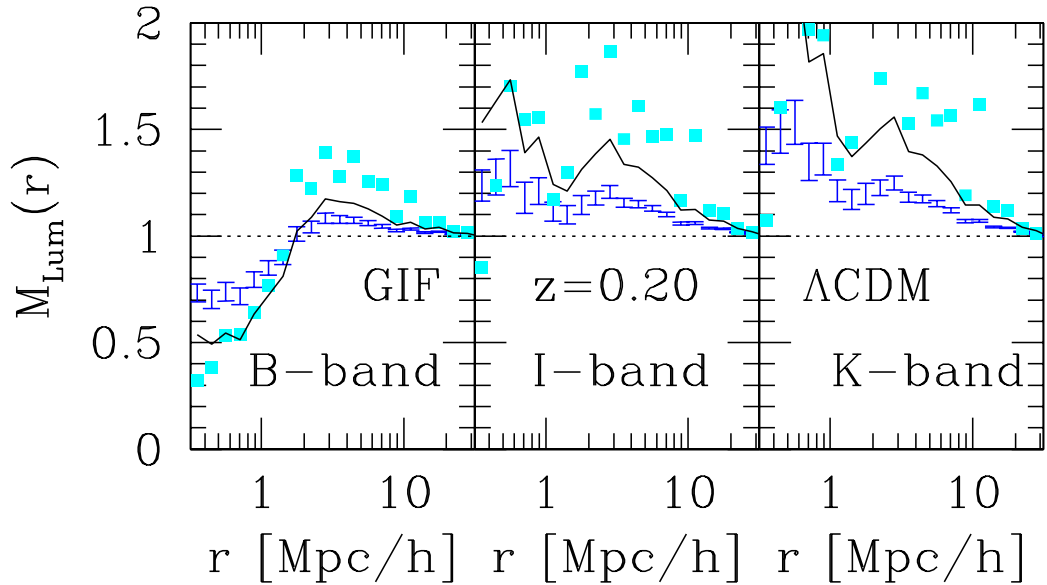}
 \vspace{-7.7cm}
 \caption{Comparison of statistic when normalized luminosity is the mark 
  (error bars) and when the square of the luminosity (suitably normalized) 
  is the mark.  Curves show that $M_L^2(r)\sim M_{L^2}(r)$, with 
  approximate equality on large scales where $M_L(r)\to 1$.  }
\label{rescaleBIK}
\end{figure*}

Figures~\ref{pdfBK} and~\ref{rescaleBIK} illustrate that the main 
features of the toy model are reproduced in the much more complex 
mark distributions studied in the main text.  Figure~\ref{pdfBK}
compares the rescaled distributions when $B$- and $K$-band 
luminosity is the mark, and when the squares of the luminosities 
are the marks.  Notice that although the distributions of normalized 
$L_B$ and $L_K$ are rather similar, the associated marked statistics 
shown in Figure~\ref{xilum} are very different.  Note also that 
the normalized distributions of $L_B^2$ and $L_K^2$ are substantially 
more skewed than those of $L_B$ and $L_K$.  

Figure~\ref{rescaleBIK} compares the marked statistics when the 
mark is $B$, $I$ or $K$-band luminosity (error bars), and when 
the mark is the square of the luminosity (filled squares).  
Note that both $M_L$ and $M_{L^2}$ show similar scale dependence, 
in agreement with the toy model discussed above.  In particular, 
note that the gross differences between the marked statistics 
when $L_B$ and $L_K$ are the marks are also obvious when $L_B^2$ 
and $L_K^2$ are used as the marks.  The jagged curves in the 
different panels show that $M_L^2$ provides a good, but by no 
means perfect, approximation for $M_{L^2}$.

We remarked in Section~\ref{norm} that monotonic rescaling of the 
marks may also be useful if one wishes to study marked statistics 
when the marks are not positive definite.  A specific example is 
galaxy color (although it happens that the distribution of $V-I$ 
in the models we have studied in the previous section is positive 
definite, so that this was not an issue).  We have studied the 
effect of rescaling from $V-I$ to $L_I/L_V \equiv 10^{-0.4(V-I)}$, 
with a curious result.  Because one of the marks is a monotonic 
(in this case exponential) transformation of the other, it is 
natural to wonder if $M_{V-I}$ or $M_{L_I/L_V}$ shows a clearer 
signal.  One might reasonably expect that the distribution of 
$V-I$ will have smaller tails than that of $L_I/L_V$.  Since 
$M_{V-I}$ increases on small scales, one might expect this 
increase to be even more pronounced for $M_{L_I/L_V}$.  On the 
other hand, since the precision of the measurement depends on 
the variance of the mark, one might expect $M_{L_I/L_V}$ to be 
the noisier measurement.  However, what matters is not the distribution 
of the mark itself, but that of the mark normalized by its mean 
value.  It happens that, upon normalizing by the mean of the rescaled 
mark, the distributions of $V-I$ and $L_I/L_V$ are almost 
indistinguishable!  Hence, the resulting marked statistics are also 
almost indistinguishable.  

It is possible to be more quantitative about this point.  
Figures~\ref{gmarks} and~\ref{rescaleBIK} show that while the 
qualitative scale dependence of $(1+W)/(1+\xi)$ does not depend on 
rescaling, it does matter quantitatively.  
This raises the question of whether or not rescaling is advisable:  
if rescaling produces a marked statistic that differs more from 
unity than did the original mark, then perhaps rescaling is desirable.  
For instance, suppose one rescales to a distribution which has longer 
tails than the original [e.g., replace all marks $w$ with $\exp(w)$].  
If $M(r)>1$ on small scales originally, then it is possible that 
after rescaling, $M(r)$ is even larger than before.  
To address whether this represents a real improvement in signal, 
one must also consider the precision with which the rescaled marked 
statistic can be measured.  Thus, the relevant question is whether 
or not this rescaling has also increased the statistical significance 
of the difference from unity.  Appendix~A provides a discussion of 
the precision with which marked statistics can be estimated.  
Equation~(\ref{ePoisson}) there shows that the ratio of the rms 
value of the marked statistic to its mean value depends on the ratio 
of $\langle w_1^2w_2^2|r\rangle$ to $\langle w_1w_2|r\rangle^2$.  
Hence, rescaling is desirable if $W^2W^2/[WW]^2$ for the rescaled 
distribution is smaller than for the original distribution, or if 
the difference between the $WW/DD$ and unity, when expressed in 
units of the rms, $[1 -DD/WW]/\sqrt{W^2W^2/[WW]^2 - 1/DD}$, is larger.  

\section{Discussion}\label{final}
Weighting galaxies by different marks (luminosity, star formation rate, 
etc.) yields datasets which are each biased differently relative to 
the underlying dark matter distribution.  Study of marked statistics 
allows one to address issues such as:  
 Which marks, if any, result in a weighting of the galaxy distribution 
 which minimizes the bias relative to the dark matter?  
 Over what redshift range is this mark the least biased?  
 Cross correlations of marks also allow one to study, for example, if 
 star formation rate, X-ray hardness or AGN activity are correlated 
 with local density; 
 if the color of a galaxy is correlated with the luminosity of its 
 neighbour (the color of a galaxy is known to be strongly correlated 
 with its own luminosity); and 
 if luminosity dependent clustering is really due to morphological 
 segregation.  

A discussion of the bias and variance of estimators of marked 
statistics is provided in Appendix~\ref{estimator}.    
We also illustrated the use of marked statistics by showing how 
they behave in semi-analytic galaxy formation models.  
The models predict that the luminosity-weighted corrrelation function 
should depend strongly on waveband (Figure~\ref{xilum}):  
close pairs should be more luminous than more widely separated pairs 
if the luminosity is measured at long wavelengths such as the 
$K$-band.  If measured using the light from shorter wavelengths 
such as the $B$-band, the trend should be reversed:  close pairs 
should be less luminous.
If the models are correct, then luminosity marked correlations from 
the HST ACS and 2MASS should resemble the bottom left and right panels, 
respectively---they should be very different from each other.  

Marked statistics also show that, in the models, close pairs of 
galaxies (separations smaller than $\sim 1$~Mpc) have redder colors 
and smaller star formation rates than more widely separated pairs 
(Figure~\ref{xisfr}), suggesting that objects in dense regions 
regions are redder and have smaller star formation rates.  
These trends are similar to the morphology-density (Dressler 1980) 
and star-formation rate-density relations (Lewis et al. 2002; 
Gomez et al. 2003).  In addition, the stellar masses of objects in 
dense regions are larger, and the ratio of the typical stellar mass 
in dense regions to the average value is larger today than it was 
in the past (Figure~\ref{ximstar}).  

We also showed that marked statistics provide simple tests of how 
correlations between observables depend on environment.  
As a specific example, the analysis of Section~\ref{correlations}, 
when combined with the measurements of $M_{\rm stars}(r)$ and 
$M_{\rm L_K}$ (Figure~\ref{ximstar}) and the fact that 
$L_K\propto M_{\rm stars}$ (Figure~\ref{m2l}) suggests that the 
mass-to-light ratio is larger in dense regions.  
It will be interesting to make similar measurements of marked 
statistics to determine whether or not other well-known correlations 
between galaxy observables correlate with environment.  
For instance, to test if the Fundamental Plane populated by 
early-type galaxy observables depends on environment, one would set 
$m=$ size and $w=$ the Fundamental Plane combination of surface 
brightness and velocity dispersion.  

The virtue of marked statistics for identifying and quantifying 
trends with environment is that they do not rely on a specific choice 
of scale on which to define the larger scale environment, nor do they 
require a subjective division of galaxy environments into cluster and 
field.  
Indeed, marked correlations are particularly well-suited to studying 
correlations between galaxy properties and their environment, when the 
correlations are weak (Sheth \& Tormen 2004).  
Because marked correlations do not pre-select a specific scale on 
which to study correlations with environment, they offer a powerful 
and complimentary approach to more traditional techniques which do 
require specific definitions of environment (e.g. Kauffmann et al. 2004; 
Hogg et al. 2005).  
Quantifying such correlations is interesting because galaxy formation 
models make specific predictions for environmental trends.  
This is because these models assume that galaxies form within dark 
matter halos, and the physical properties of galaxies are determined 
by the halos in which they form (White \& Rees 1978; White \& Frenk 1991).  
Therefore, the statistical properties of a given galaxy population 
are determined by the properties of the parent halo population.  

The galaxy formation models of Kauffmann et al. (1999) and 
Cole et al. (2000) are based on this paradigm:  they make rather 
similar assumptions for the complicated gastrophysics of galaxy 
formation.  However, they make different assumptions about the 
resulting correlations with environment.  
In particular, because Kauffmann et al. (1999) use halo merger history 
trees from the simulations, their models include correlations between 
halo formation histories and environment.  In contrast, 
Cole et al. (2000) use Monte Carlo merger histories, which explicitly 
ignore any correlation between halo formation and environment.  
Marked correlation functions provide an efficient way of quantifying 
the differences which result.  
In this regard, Skibba, Sheth \& Connolly (2006) show how the observed 
luminosity dependence of clustering, when combined with a measurement 
of the luminosity weighted correlation function allows a sensitive test 
of the environmental dependence of galaxy clustering.  

There are two main disadvantages associated with using marked 
statistics.  First, although considerable effort has been spent in 
developing accurate models of how the galaxy correlation function 
depends on scale and time (see Cooray \& Sheth 2002 for a review), 
there has been little parallel development of how marked correlations 
are expected to depend on scale and time.  Hence, interpretting the 
measurements is not straightforward.  However, this is beginning to 
change:  Sheth, Abbas \& Skibba (2004) show how to describe the weighted 
correlation function using the halo-model of large-scale structure, 
when galaxies are associated with halo substructure, and Sheth (2005) 
shows how the halo model can be used to describe all the marked 
statistics described here.  

\begin{figure}
\centering 
 \vspace{-1cm}
 \includegraphics[width=1.1\hsize]{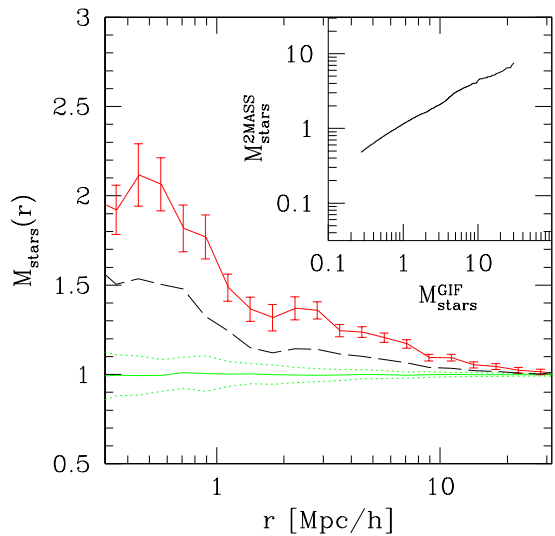}
 \vspace{-1cm}
 \caption{Comparison of statistic when normalized stellar mass is the 
          mark.  Top curve with error bars shows the result when the 
          stellar mass is that from the GIF galaxy formation simulation, 
          whereas dashed curve shows the statistic after applying a 
          monotonic rescaling to the marks (shown as the inset in the 
          upper right corner), so that they match the distribution of 
          stellar masses reported by Cole et al. (2001).  }
\label{rescaled2dF}
\end{figure}

Second, we have assumed that all analysis of marked statistics will 
be performed on volume limited catalogs.  While direct measurement 
(using the estimator discussed here) of marked statistics in 
magnitude limited surveys is straightforward, interpretation is 
more complicated.  For instance, one might use the halo-model 
framework to model the measurement.  Development of a measurement 
algorithm which accounts for the effects of the magnitude limit is 
the subject of work in progress.  There is, in addition, the 
inconvenience associated with the fact that our analysis assumes 
that the marks are positive definite, but arguments in 
Section~\ref{rescale} suggest that monotonic rescaling to a 
distribution which is positive definite will still yield useful 
results.  Of course, all this assumes that the marks themselves 
have been accurately and precisely determined from the data.  

Marked statistics are sensitive to the distribution of marks 
in the dataset.  Because it is likely that the semi-analytic models 
will not match the observed distribution of galaxy properties 
(e.g., the current generation of models does not match the SDSS 
luminosity function, and the distribution of stellar masses does 
not match that infered from 2MASS), we envision that constraints 
on models will take the following form.  
If the models successfully describe the luminosity dependence of 
clustering, then they have successfully identified the halos in which 
galaxies form.  Because the marked correlations are not changed in 
any essential way if the absolute values of the weights are altered 
without changing the rank ordering (Section~\ref{rescale}), 
marked statistics provide a way of determining if, in addition to 
correctly identifying the halos in which galaxies form, the models 
also determine the correct rank ordering of galaxy properties within 
halos, if not their absolute values.  

Figure~\ref{rescaled2dF} illustrates this procedure. 
It shows $(1+W)/(1+\xi)$ when stellar mass is the mark:  
the top curve with error bars shows the measurement in the GIF 
simulations (i.e., it is the same as in Figure~\ref{ximstar}), 
whereas the bottom dashed curve shows the result of rescaling the 
stellar masses so that they match the distribution reported by 
Cole et al. (2001) from their analysis of 2dF and 2MASS.  To do 
this, we assumed that the marks in the GIF simulation are for the 
most massive galaxies in the simulation box.  We then matched 
cumulative number densities, meaning that we worked our way down the 
GIF and Cole et al. stellar mass functions, assigning new masses to 
the semi-analytic galaxies until all 14665 objects had been assigned 
a 2dF-2MASS mass.  Finally, we normalized these rescaled masses by 
their mean.  The original and rescaled marks, normalized by their 
mean values, are shown in the inset in the top right corner of the 
Figure.  This shows that, after rescaling, the marks span a smaller 
range of values around their mean value than they did originally.  
This has the effect of decreasing the amplitude of the rescaled 
marked statistic, but does not change the qualitative trend for close 
pairs to be more massive.  More importantly, this illustrates that 
if the distribution of marks in the models does not match the data, 
then the marked statistic will also be discrepant.  

The measurements presented here serve as benchmarks against which 
to compare measurements from data such as the SDSS (Abazajian et al. 2005).  
Measurements in the SDSS database using color, velocity dispersion, 
effective radius, morphology, star formation rate indicators, local 
density, and so on, as marks are underway (Skibba, Sheth \& Connolly 2006).  
Comparison with predictions such as those described above will 
constrain the galaxy formation models.  

\section*{acknowledgments}
We would like to thank the Virgo Consortium for making their 
simulations available to the public, the Pittsburgh Computational 
Astrophysics group (PiCA) for developing fast algorithms with which 
to measure correlation functions, and the referee for a detailed 
and helpful report.  
This work is supported by NASA and the NSF under grants NAG5-13270 and 
AST-0307747 and AST-0520647.

\appendix

\section{Error estimates}\label{estimator}
In what follows, we provide error estimates of the marked statistic 
most studied in the main text, the ratio of the weighted and unweighted 
correlation functions: $(1+W)/(1+\xi)$.  We do this in two steps;
the first assumes the marks are perfectly measured, and the second 
includes the effects of measurement errors.  In either case, the 
analysis assumes that the mean mark can be reliably determined from 
the data.  Our analysis follows that of Peebles (1980), 
Mo, Jing \& B\"orner (1992), Landy \& Szalay (1993) and 
Bernstein (1994).

\subsection{An unbiased estimator}
We estimate the statistic as $WW/DD$, where $WW$ is the sum over all 
pairs with separation $r$, weighting each member of the pair by its 
mark, and $DD$ is the total number of such pairs.  In what follows, 
we assume that the marks have all been normalized by the mean mark, 
so that $DD$ really is the total number of pairs.  (We discuss an 
alternative procedure, in which $DD$ is replaced by $W_sW_s$, which 
denotes the weighted pair counts in which the weights have been 
scrambled, shortly.)  
Let $\langle WW\rangle$ and $\langle DD\rangle$ denote the mean 
values of these statistics, where the mean is over an ensemble of 
realizations of the point process.  The counts in any particular 
realization differ from this ensemble average value, so that 
\begin{equation}
 \Bigl\langle{WW\over DD}\Bigr\rangle 
   = {\langle WW\rangle\over \langle DD\rangle} 
              \Bigl\langle {1+\omega\over 1+\delta}\Bigl\rangle
   \approx {\langle WW\rangle\over \langle DD\rangle}\,
       \Bigl\langle 1 + \omega - \delta - \omega\delta + \delta^2\Bigr\rangle.
\end{equation}
The final expression follows from assuming $\delta\ll 1$ and keeping 
terms to second order.  Now, $\langle\omega\rangle=0$ and 
$\langle\delta\rangle=0$ by definition, so the average of the ratio 
(the left hand side) equals the ratio of the averages (the right hand 
side) if $\langle\omega\delta\rangle = \langle\delta^2\rangle$.  
We can estimate these terms as follows.  Let 
\begin{equation}
 \Bigl\langle DD\Bigr\rangle 
   = \sum_{i<j} n_i n_j\,\Theta_{ij}(r) = U_2(r) 
\end{equation}
and
\begin{equation}
 \Bigl\langle WW\Bigr\rangle 
   = \sum_{i<j} w_i w_j\,\Theta_{ij}(r) 
   = \Bigl\langle w_1w_2|r\Bigr\rangle\, U_2(r),
\end{equation}
where $U_2(r)$ is proportional to the total number of pairs, 
$N(N-1)/2$, times a geometrical factor, $G_2(r)$, which represents 
the fraction of these pairs which have separation $r$, given the 
survey geometry (see, e.g., Landy \& Szalay 1993).  Then 
\begin{eqnarray}
 {\langle{DD\cdot DD}\rangle \over \langle DD\rangle \langle DD\rangle}
   &=& 1 + \Bigl\langle\delta^2\Bigr\rangle \nonumber \\
   &=& {\sum_{i<j} n_i n_j\,\Theta_{ij}(r) \sum_{k<l} n_k n_l\,\Theta_{kl}(r)
            \over U_2(r)^2} \nonumber\\
   &=& {U_4(r) + U_3(r) + U_2(r)\over U_2(r)^2},
\end{eqnarray}
where $U_4(r) \equiv N(N-1)(N-2)(N-3)/4 \,G^2_2(r)$, and 
$U_3(r) \equiv N(N-1)(N-2)\,G_3(r)$ are products of the total number 
of quadruples and triples times geometrical factors, defined similarly 
to $U_2(r)$.  
Similarly, 
\begin{eqnarray}
 {\langle{WW\cdot DD}\rangle \over \langle WW\rangle \langle DD\rangle}
   &=& 1 + \Bigl\langle\omega\delta\Bigr\rangle \nonumber \\
   &=& {\sum_{i<j} w_i w_j\,\Theta_{ij}(r) \sum_{k<l} n_k n_l\,\Theta_{kl}(r)
            \over \langle w_1 w_2|r\rangle\,U_2(r)^2} \nonumber\\
   &=& {\langle w_1 w_2|r\rangle\, U_4(r) 
              + \langle w_1 w_2|r\rangle\, U_3(r) 
              + \langle w_1 w_2|r\rangle\,U_2(r)
             \over \langle w_1 w_2|r\rangle\,U_2(r)^2}\nonumber \\  
   &=& 1 + \Bigl\langle\delta^2\Bigr\rangle.
\end{eqnarray}
Thus, 
\begin{equation}
 \Bigl\langle{WW\over DD}\Bigr\rangle 
   = {\langle WW\rangle\over \langle DD\rangle};
\end{equation}
this estimator is unbiased.  

Notice that if we had not normalized the marks by their mean value, 
but had replaced $DD$ by $W_sW_s$, then the estimator would be unbiased 
provided we treated each of the $W_s$ terms as being drawn from a 
different realization of the scrambled marks.  This is because terms 
of the form $\langle w^2\rangle$ in the expression for 
$\langle W_sW_s\cdot W_sW_s\rangle$ lead to a bias.  
Thus, the weights in the numerator of $WW/W_sW_s$ must be treated 
differently from those in the denominator, making the notation 
somewhat confusing.  This is why we prefer to use $WW/DD$ (with 
the understanding that the weights have been normalized by their 
mean value), rather than $WW/W_sW_s$ as our estimator.  

\subsection{Variance within a bin}
The variance of the estimator is 
\begin{eqnarray}
 {\rm Var}\left[{WW\over DD}\right]
    &=& {\langle WW\rangle^2\over \langle DD\rangle^2} 
       \Bigl\langle {(1+\omega)^2\over (1+\delta)^2}\Bigl\rangle 
       - \Bigl\langle {WW\over DD}\Bigr\rangle^2\nonumber\\
    &=& {\langle WW\rangle^2\over \langle DD\rangle^2} 
        \Bigl\langle 1 + \omega^2 - 4\omega\delta + 3\delta^2\Bigr\rangle 
       - \Bigl\langle {WW\over DD}\Bigr\rangle^2\nonumber\\
    &=& {\langle WW\rangle^2\over \langle DD\rangle^2}\, 
        \Bigl\langle \omega^2 - \delta^2\Bigr\rangle 
\end{eqnarray}
where the final expression uses the fact that 
 $\langle\omega\delta\rangle = \langle\delta^2\rangle$.  

The sum in quadrature of the variances of $WW$ and $DD$ is proportional 
to $\langle\omega^2\rangle + \langle\delta^2\rangle$.  
Notice that the expression above scales as the difference of these 
two terms rather than than the sum, so it can be {\em substantially} 
smaller.  This reflects the fact that, if $WW$ for a given realization 
of the point process is larger than the ensemble mean, it may simply 
be large because $DD$ is also large (i.e., there were more pairs).  
As a result, fluctuations in $WW$ and $DD$ are correlated.  
The estimate of the true variance above accounts for this correlation.  

To see what it is, note that 
\begin{eqnarray}
 \!&& {\langle{WW\cdot WW}\rangle \over \langle WW\rangle \langle WW\rangle}
   = 1 + \Bigl\langle\omega^2\Bigr\rangle \nonumber \\
   &=& {\sum_{i<j} w_i w_j\,\Theta_{ij}(r) \sum_{k<l} w_k w_l\,\Theta_{kl}(r)
            \over \langle w_1 w_2|r\rangle^2\,U_2(r)^2} \nonumber\\
   &\approx& {U_4(r)\over U^2_2(r)} +
              {\langle w_1^2 w_2 w_3|r\rangle\, U_3(r) 
              + \langle w_1^2 w_2^2|r\rangle\,U_2(r)
             \over \langle w_1 w_2|r\rangle^2\,U_2(r)^2}.  
\end{eqnarray}
so the ratio of the variance to the square of the mean is 
\begin{eqnarray}
 {{\rm Var}[WW/DD]\over \langle WW/DD\rangle^2}
   &\approx& {\langle w_1^2 w_2 w_3|r\rangle - \langle w_1 w_2|r\rangle^2
         \over \langle w_1 w_2|r\rangle^2}\, {U_3(r)\over U^2_2(r)}\nonumber\\
   && + {\langle w_1^2 w_2^2|r\rangle - \langle w_1 w_2|r\rangle^2
            \over \langle w_1 w_2|r\rangle^2}\,{U_2(r)\over U_2(r)^2} .
\end{eqnarray}
On large scales, where correlations tend to be weak, 
 $U_3\approx N_{\rm gal}\,(\bar n\, {\rm d}V)^2 (1+\xi)^3/2$ 
and 
 $U_2\approx N_{\rm gal}\,\bar n\, {\rm d}V\,(1+\xi)/2$ 
so 
 $U_3/U_2^2 \approx 2(1+\xi)/N_{\rm gal}$.  
Hence, the first term in the expression above scales as $1/N_{\rm gal}$.  
The second term scales inversely with the number of pairs, so, for 
a sufficiently large survey, the first term dominates on large scales, 
whereas the second term dominates on small scales.  


The second term scales sensibly as the inverse of the pair counts, 
times a term which resembles a variance of mark pairs.  
It is useful to think of this term as 
\begin{displaymath}
 \left[{\langle w_1^2 w_2^2|r\rangle\over \langle w_1 w_2|r\rangle^2} -1\right]
 \Bigl/U_2(r)
   = \left[{M_{w^2}(r)\over M^2_w(r)}{\langle w^2\rangle^2\over 
           \langle w\rangle^4} -1\right] \Bigl/U_2(r);
 \label{ePoisson}
\end{displaymath}
here $M_{w^2}(r)$ denotes the statistic when the mark is the square 
of the weight, rather than the weight itself.  Thus, this term 
can be estimated directly from the data as  $W^2W^2/[WW]^2 - 1/DD$.
Notice that if $M_{w^2}= M_w^2$, then this term can be written in 
terms of the variances of $w$ and $w^2$, so the only scale dependence  
comes from the factor of $U_2$.  Figure~\ref{rescaleBIK} in the main 
text shows that this approximation may not be wildly-off in 
astrophysical datasets.  

The first term is more problematic; its presence suggests that a 
better estimator than $WW/DD$ for our marked statistic can be found, 
just as a better estimator than $DD/RR$ can be found for the 
unweighted statistic $1+\xi$.  Writing down this alternative 
estimator is beyond the scope of the present paper.   However, to see 
the effect of this extra term, suppose that there are no correlations, 
either between the points, or between the marks.  
This is not an unreasonable assumption on the large scales where the 
first term is expected to dominate.  In this limit, 
\begin{equation}
 {{\rm Var}[WW/DD]\over \langle WW/DD\rangle^2}
   \approx \left[{\langle w^2\rangle\over \langle w\rangle^2}-1\right]
         {2\over N_{\rm gal}}.
\end{equation}
The form of this expression suggests that the variance should be 
well approximated by adding 
\begin{equation}
   \left[{\langle w^2|r\rangle\over \langle w_1w_2|r\rangle}-1\right]
         {2\,[1+\xi(r)]\over N_{\rm gal}} 
 \label{nonPoisson}
\end{equation}
to $W^2W^2/[WW]^2 - 1/DD$.  Figure~\ref{errors} in the main text 
illustrates how these two terms are expected to contribute to the 
error budget in typical astrophysical datasets.  

The analysis so far does not account for the possibility that, 
if the dataset is small, then the estimated pair counts are not 
from a fair sample, so they will be biased.  This introduces 
additional terms into the error budget, which can be estimated by 
relatively straightforward but tedious extension of the analysis in 
Landy \& Szalay (1993).  
Also absent from this error budget is the fact that small datasets 
may not provide accurate measurements of the mean mark.  Since our 
analysis assumes that all marks have been normalized by this mean 
measured in the dataset, this introduces an additional source of 
error.  A more detailed account of these two additional sources of 
error will be presented elsewhere.  Note that astrophysical datasets 
are now becoming sufficiently large that small sample issues are less 
likely to be a serious concern, so the estimates here should be 
reasonably accurate.

\subsection{Covariance between bins}
In this case, we are interested in 
\begin{eqnarray}
{\rm cov}\left[{W_1W_1\over D_1D_1}{W_2W_2\over D_2D_2}\right] 
  &=&  {\langle W_1W_1\rangle\langle W_2W_2\rangle\over 
        \langle D_1D_1\rangle\langle D_2D_2\rangle} \,
       \Bigl\langle \delta_1\delta_2 + \omega_1\omega_2 \nonumber\\
  &&  \qquad\qquad     - \omega_1\delta_2 - \omega_2\delta_1\Bigr\rangle
\end{eqnarray}
where we have used the fact that 
\begin{eqnarray*}
 \left\langle{W_1W_1\over D_1D_1}{W_2W_2\over D_2D_2}\right\rangle 
  &=&  {\langle W_1W_1\rangle\langle W_2W_2\rangle\over 
        \langle D_1D_1\rangle\langle D_2D_2\rangle} \,
       \Bigl\langle 1 + \delta_1^2 + \delta_2^2 + \delta_1\delta_2 \nonumber\\
  &&            - \omega_1\delta_1 - \omega_2\delta_2
                + \omega_1\omega_2 
                - \omega_1\delta_2 - \omega_2\delta_1\Bigr\rangle
\end{eqnarray*}
and we have set $\langle\omega_i\delta_i\rangle = \langle\delta_i^2\rangle$.  
Hence, the covariance depends on 
\begin{equation}
 {\langle D_iD_i\cdot D_jD_j\rangle\over 
  \langle D_iD_i\rangle \langle D_jD_j\rangle} 
  = 1 + \Bigl\langle \delta_i\delta_j\Bigr\rangle 
  = 1 + {U_3(r_i,r_j)\over U_2(r_i)\,U_2(r_j)},
\end{equation}
\begin{equation}
 {\langle W_iW_i\cdot D_jD_j\rangle\over 
  \langle W_iW_i\rangle \langle D_jD_j\rangle} 
  = 1 + \Bigl\langle w_i\delta_j\Bigr\rangle
  = 1 + \Bigl\langle\delta_i\delta_j\Bigr\rangle,
\end{equation}
and
\begin{eqnarray}
 {\langle W_iW_i\cdot W_jW_j\rangle\over 
  \langle W_iW_i\rangle \langle W_jW_j\rangle} 
  &=& 1 + \Bigl\langle w_iw_j\Bigr\rangle \nonumber\\
  &=& 1 + W_3(r_i,r_j)\,{U_3(r_i,r_j)\over U_2(r_i)\,U_2(r_j)},
\end{eqnarray}
with 
\begin{equation}
 W_3(r_i,r_j) = {\langle w_1^2w_2w_3|r_i,r_j\rangle\over
                 \langle w_1w_2|r_i\rangle\langle w_1w_3|r_j\rangle} - 1.
\end{equation}

\subsection{Measurement errors}
The analysis above assumes the marks are perfectly measured.  
If, instead, the marks are imprecisely measured, but, while 
imprecise, are unbiased, and if the measurement error is not 
correlated with spatial position, then the previous expressions 
hold with 
$\langle\omega^2\rangle\to\langle\omega^2+\epsilon^2\rangle$ 
where $\langle\epsilon^2\rangle$ denotes the variance of the 
error on mark.  Hence, the measurement errors do not bias the 
marked statistic, but they do increase Var$[WW/DD]$, thus 
decreasing the precision of the measurement.  
Of course, if the measurement of the marks themselves is 
systematically biased, then the marked statistics will also be 
biased.

\label{lastpage}


\begin{thebibliography}{9}

\bibitem{dr3} Abazajian K., et al., 2005, AJ, 129, 1755

\bibitem{BK00} Beisbart C., Kerscher M., 2000, ApJ, 545, 6

\bibitem{BC99} Benoist C., Cappi A., da Costa L. N., Maurogordato S., 
           Bouchet F., Schaeffer R., 1999, ApJ, 514, 563

\bibitem{B94} Bernstein G., 1994, ApJ, 424, 569

\bibitem{2dF2Mass} Cole S., Norberg, P, Baugh C. M., Frenk C. S., et al., 
                 2001, MNRAS, 326, 255

\bibitem{Durham} Cole S., Lacey C., Baugh C. M., Frenk C. S., 
                   2000, MNRAS, 321, 559

\bibitem{2dF01} Colless M., et al., 2001, MNRAS, 328, 1039

\bibitem{CS95} Connolly A. J., Szalay A. S., Bershady M. A., Kinney A. L.,
           Calzetti D., 1995, AJ, 110(3), 1071

\bibitem{CSJ02} Connolly A. J., Scranton R., Johnston D., et al., 
                2002, ApJ, 579, 42

\bibitem{CS02} Cooray A., Sheth R. K., 2002, Phys. Rep., 372, 1

\bibitem{AD80} Dressler A., 1980, ApJ, 236, 351

\bibitem{HV} Evrard A. E., MacFarland T. J., Couchman H. M. P., 
           Colberg J. M, et al., 2002, ApJ, 573, 7

\bibitem{FJ76} Faber S. M., Jackson R. E., 1976, ApJ, 204, 668

\bibitem{sdssSFR} Gomez P., et al., 2003, ApJ, 584, 210

\bibitem{AJSH88} Hamilton A. J. S., 1988, ApJL, 331, L59

\bibitem{PSCz} A. J. S. Hamilton,  M. Tegmark, 2002, MNRAS, 330, 506

\bibitem{GIF} Kauffmann G. A. M., Colberg J. M., Diaferio A., White S. D. M., 
           1999, MNRAS, 303, 188

\bibitem{K04} Kauffmann G. A. M., White S. D. M., Heckman T. M., 
           M\'enard B., Brinchmann J., Charlot S., Tremonti C., Brinkmann J.,
           2004, MNRAS, 353, 713

\bibitem{NN96} Lahav O., Naim A., Sodr\'e L., Storrie-Lombardi M. C., 1996, 
           MNRAS, 283, 207

\bibitem{LS93} Landy S. D., Szalay A. S., 1993, ApJ, 412, 64

\bibitem{2dFsfr} Lewis I., et al., 2002, MNRAS, 334, 673

\bibitem{APM} Maddox S. J., Efstathiou G., Sutherland W. J., Loveday J., 1990, 
           MNRAS, 242, 43P

\bibitem{MBZ89} Mo H. J., B\"orner G., Zhou Y., 1989, AA, 221, 191

\bibitem{MJB92} Mo H. J., Jing Y. P., B\"orner G., 1992, ApJ, 392, 452

\bibitem{MW96} Mo H. J., White S. D. M., 1996, MNRAS, 282, 347

\bibitem{PiCA} Moore A., Connolly A. J., Genovese C., Gray A., Grone L.,
           Kanidoris N., Nichol R. C., Schneider J., Szalay A. S., 
           Szapudi I., Wasserman L., 
           in Mining the Sky, Proceedings of the MPA/ESO/MPE Workshop.  
           Eds. A. J. Banday, S. Zaroubi \& M. Bartelmann.  
           Heidelberg:  Springer Verlag, p.71 (2001)

\bibitem{PN02} Norberg P., Baugh C. M., Hawkins E., Maddox S. J., et al., 
           2002, MNRAS, 332, 827

\bibitem{2dFPk} Percival W. P., et al., 2001, MNRAS, 327, 1297

\bibitem{LSS80} P. J. E. Peebles (1980), The Large Scale Structure of the Universe. Princeton.  

\bibitem{RKS05} Sheth R. K., 2005, MNRAS, ???, ???

\bibitem{SL99} Sheth R. K., Lemson G., 1999, MNRAS, 304, 767

\bibitem{SAS03} Sheth R.K., Abbas U., Skibba R.A., in Diaferio A., ed, 
           Proc. IAU Coll. 195, 
           Outskirts of galaxy clusters: intense life in the suburbs, 
           CUP, Cambridge, p.~349

\bibitem{SSC} Skibba R., Sheth R. K., Connolly A. J., 2006, MNRAS, 
           submitted (astro-ph/0512???)
 
\bibitem{SP99} Somerville R. S., Primack J. R., 1999, MNRAS, 310, 1087

\bibitem{Millennium} Springel V., et al., 2005, Nature, 435, 629
 
\bibitem{NN92} Storrie-Lombardi M. S., Lahav O., Sodr\'e L., 
           Storrie-Lombardi L. J., 1992, MNRAS, 259, 8P

\bibitem{stoyan} Stoyan D., 1984, Math. Nachr., 116, 197

\bibitem{stoyan2} Stoyan D., Stoyan H., 1994, 
       Fractals, Random Shapes, and Point Fields. Wiley \& Sons, Chichester

\bibitem{TK69} Totsuji H., Kihara T., 1969, PASJ, 21, 221

\bibitem{TF77} Tully R. B., Fisher J. R., 1977, A\&A, 54(3), 661

\bibitem{WR78} White S. D. M., Rees M. J., 1978, MNRAS, 183, 341

\bibitem{WF91} White S. D. M., Frenk C. S., 1991, ApJ, 379, 52

\bibitem{WCP98} Willmer C. N. A., da Costa L. N., Pellegrini P. S., 1998, AJ, 115(3), 869


\bibitem{IZ02} Zehavi I., Blanton M. R., Frieman J. A., 
               Weinberg D. H., et al., 2002, ApJ, 571, 172

\bibitem{IZ05} Zehavi I., Blanton M. R., Frieman J. A., 
               Weinberg D. H., et al., 2005, ApJ, 630, 1

\end{thebibliography}
\end{document}